# Long-period events with strikingly regular temporal patterns on Katla volcano's south flank (Iceland)

Giulia Sgattoni[1,2,3]*, Zeinab Jeddi[3,4], Ólafur Guðmundsson[3], Páll Einarsson[2], Ari Tryggvason[3], Björn Lund[3], Federico Lucchi[1]

[1] *Department of Biological, Geological and Environmental Sciences, University of Bologna, Bologna, Italy*

[2] *Institute of Earth Sciences, Science Institute, University of Iceland, Reykjavik, Iceland*

[3] *Department of Earth Sciences, Uppsala University, Uppsala, Sweden*

[4] *CNDS, Centre for Natural Disaster Science, Uppsala University, Uppsala, Sweden*

*\*Corresponding author:* giulia.sgattoni2@unibo.it

## Abstract

Katla is a threatening volcano in Iceland, partly covered by the Mýrdalsjökull ice cap. The volcano has a large caldera with several active geothermal areas. A peculiar cluster of long-period seismic events started on Katla's south flank in July 2011, during an unrest episode in the caldera that culminated in a glacier outburst. The seismic events were tightly clustered at shallow depth in the Gvendarfell area, 4 km south of the caldera, under a small glacier stream on the southern margin of Mýrdalsjökull. No seismic events were known to have occurred in this area before. The most striking feature of this seismic cluster is its temporal pattern, characterized by regular intervals between repeating seismic events, modulated by a seasonal variation. Remarkable is also the stability of both the time and waveform features over a long time period, around 3.5 years. No comparable examples have been found in the literature. Both volcanic and glacial processes can produce similar waveforms and therefore have to be considered as potential seismic sources. Discerning between these two causes is critical for monitoring glacier-clad volcanoes and has been controversial at Katla. For this new seismic cluster on the south flank we regard volcano-related processes as more likely than glacial ones for the following reasons: 1) the seismic activity started during an unrest episode involving sudden melting of the glacier and a jökulhlaup; 2) the glacier stream is small and stagnant; 3) the seismicity remains regular and stable for years; 4) there is no apparent correlation with short-term weather changes, such as rain storms. We suggest that a small, shallow hydrothermal system was activated on Katla's south flank in 2011, either by a minor magmatic injection or by changes of permeability in a local crack system.

# 1. Introduction

Katla is one of the most active and threatening volcanoes in Iceland. Partly covered by Mýrdalsjökull glacier, its volcanic activity has been dominated by large explosive phreatomagmatic eruptions, the last occurring in 1918. The current repose time is the longest known in history (Larsen, 2000). Katla is located just east of Eyjafjallajökull and the two volcanoes appear to be tectonically connected (Einarsson and Brandsdóttir, 2000). Several cases are known in history when Katla's eruption followed its neighbour's. For this reason, the 2010 Eyjafjallajökull eruption prompted scientists' concerns about a possible imminent Katla's eruption and the seismic network around the volcano was densified.

Seismic events at volcanoes are highly variable in terms of waveforms and temporal sequence evolution. They can be generated by multiple, geothermal, volcanic or tectonic processes. LP events (Long-Period; Chouet, 1996) are of particular interest as they often accompany or precede volcanic eruptions, but they are still not well understood (Chouet, 2003; McNutt, 2005). They can be related to a number of different magmatic and hydrothermal processes, usually associated with fluid movements inside the volcano. Recent studies offer new modelling of LP events as slow-rupture failure in unconsolidated volcanic materials (Bean et al., 2014).

As Katla is largely covered by Mýrdalsjökull glacier, extending over a ~600 km$^2$ area and entirely covering the summit caldera, direct field studies are not feasible and indirect methods such as seismic data analysis are important to study and monitor the volcano. Katla offers a wide variety of seismic signals to obtain insight into the volcano's internal dynamics. In this respect, it is an unusual volcano in Iceland because of its anomalously high and persistent seismicity (even during volcanic quiescence) despite its tectonic location, out of the main deformation zones. The only other Icelandic volcanoes comparable to Katla with respect to intense seismicity are located along the plate boundaries (Hengill and Bárðarbunga volcanoes; Einarsson, 1991; Jakobsdóttir, 2008).

Most of Katla's seismicity, consisting of VT (Volcano-Tectonic; Chouet, 2003; McNutt, 2005) and LP events, has primarily been recorded in two distinct source areas (e.g. Sturkell et al., 2010): i) within the summit caldera and ii) in a region in western Mýrdalsjökull named Goðabunga (Fig. 1). In July 2011, though, this general pattern changed. An increase in seismicity and a 23 hour tremor burst were recorded on July 8-9$^{th}$ 2011, associated with deepening of ice cauldrons on the ice cap and flooding from the south east rim of Mýrdalsjökull glacier that destroyed a bridge on the main road. At the same time, a new cluster of LP seismic events was detected on the south flank, just west of the Gvendarfell rise, around 4 km south of the caldera rim. Seismic events in this area of Katla had never been recorded before and they are of major interest when assessing Katla's volcanic hazard, as the south coast of Iceland is a farming area and a popular tourist destination.

At subglacial volcanoes, seismic events associated with glaciers (such as basal slip, ice falls, crevassing) and shallow low-frequency earthquakes associated with volcanic activity can produce similar waveforms (Métaxian et al., 2003; West et al., 2010; Thelen et al., 2013). The ability to distinguish between such sources is critical for monitoring glacier-clad volcanoes. In this respect, Katla aroused controversial interpretations of its seismicity, for its subglacial location and for the unusually frequent occurrence of LP seismicity during periods of volcanic quiescence (Soosalu et al., 2006; Jónsdóttir et al., 2009).

Although the LP events located on Katla's south flank in some aspects resemble those of other volcanoes, they show some peculiar characteristics, especially in their time evolution, which we have not found described in the literature. These are the subjects of the paper herein. The improved seismic network operating around Katla during 2011-2013 provided a good dataset for this purpose. Moreover, in order to improve the location estimates of these events further, a temporal deployment of three stations was done in 2014 in the epicentral area, in order to better define the absolute location of the cluster. We used cross-correlation methods to improve the event detection, thus highlighting striking time-sequence features, and a probabilistic non-linear method for the absolute location of the cluster. We also attempted to retrieve the focal mechanisms from first motion polarities.

## 2. Study area

The Katla volcanic system is located just south of the intersection between the East Volcanic Zone and the transform boundary of the South Iceland Seismic Zone (Sturkell et al., 2008). It consists of a central volcano with a 110 km$^2$ summit caldera (up to 14-km wide; Fig. 1) covered by Mýrdalsjökull glacier (Björnsson et al., 2000) and the Eldgjá fissure system which extends 75 km to the northeast (Larsen, 2000; Thordarson et al., 2001; Fig. 1). The central volcano's activity is dominated by explosive phreatomagmatic eruptions due to magma-ice interaction. Several ice cauldrons (at least 16) are located within and at the caldera rim, representing the surface expression of subglacial geothermal activity. Changes in their geometry are monitored to detect variations of geothermal heat release (Guðmundsson et al., 2007).

Seismic undershooting within the Katla caldera has revealed a zone where P wave velocities are reduced and S waves are absent; this anomaly is interpreted as evidence of a magma chamber (Guðmundsson et al., 1994). Moreover, results from an aeromagnetic survey indicate the presence of a non-magnetic body within the region of the postulated magma chamber (Jónsson and Kristjánsson, 2000). This is supported by geobarometry analyses on historical tephra samples, conducted by Budd et al. (2014), but questioned by tephra stratigraphy studies by Óladóttir et al. (2008).

Since the first sensitive seismographs were installed (in the 1960s), seismic activity has been observed in two distinct main areas: within the caldera and at Goðabunga on the western flank (Einarsson and Brandsdóttir, 2000). The seismicity inside the caldera consists of high frequency and hybrid events, probably associated with the subglacial geothermal activity (Sturkell et al., 2010) and volcano-tectonic processes. The Goðabunga cluster consists mainly of long-period shallow events and has a controversial interpretation, as a response to a slowly-rising viscous crypto-dome (Soosalu et al., 2006) or in association with ice fall events (Jónsdóttir et al., 2009). Katla's seismicity shows a seasonal variation which is stronger at Goðabunga, where the peak occurs in autumn. A less pronounced peak of activity in the caldera occurs instead during the summer (July/August; Jónsdóttir et al., 2007). This seasonal correlation has been interpreted either as a result of ice-load change and resulting pore-pressure change at the base of the glacier (Einarsson and Bransdóttir, 2000) or as enhanced glacial motion during periods of distributed subglacial water channels (Jónsdóttir et al., 2009).

Although no visible eruptions have occurred after 1918, evidence of unrest was observed in 1955, 1999 and 2011, possibly associated with minor subglacial eruptions. In 1955 this probably took place at the eastern rim of the caldera: two shallow ice-cauldrons formed and a small jökulhlaup drained from Kötlujökull, south-east Mýrdalsjökull (Thorarinsson, 1975; Rist, 1967). A similar event took place in July 1999: the seismic stations around the glacier recorded earthquakes and bursts of tremor that culminated in the release of a jökulhlaup from Sólheimajökull, south-west Mýrdalsjökull (Sigurðsson et al., 2000; Roberts et al., 2003). A new cauldron also formed on the surface of the glacier (Guðmundsson et al., 2007).

From 1999 to 2004, GPS measurements on nunataks exposed on the caldera edge revealed steady uplift of the volcano, interpreted to result from 0.01 km³ magma accumulation (Sturkell et al., 2006; 2008). A recent study by Spaans et al. (2015), suggested instead that the uplift may be due to glacial isostatic adjustment as a consequence of mass loss of Iceland's ice caps.

Guðmundsson et al. (2007) showed that increased geothermal heat output occurred in 2001-2003, based on the evolution of ice cauldrons. As this was also accompanied by greatly increased seismicity and ground uplift, they interpreted these phenomena as a result of magma accumulation and consequent dilation of the edifice leading to enhanced permeability and increased geothermal activity, in accordance with Sturkell et al. (2006).

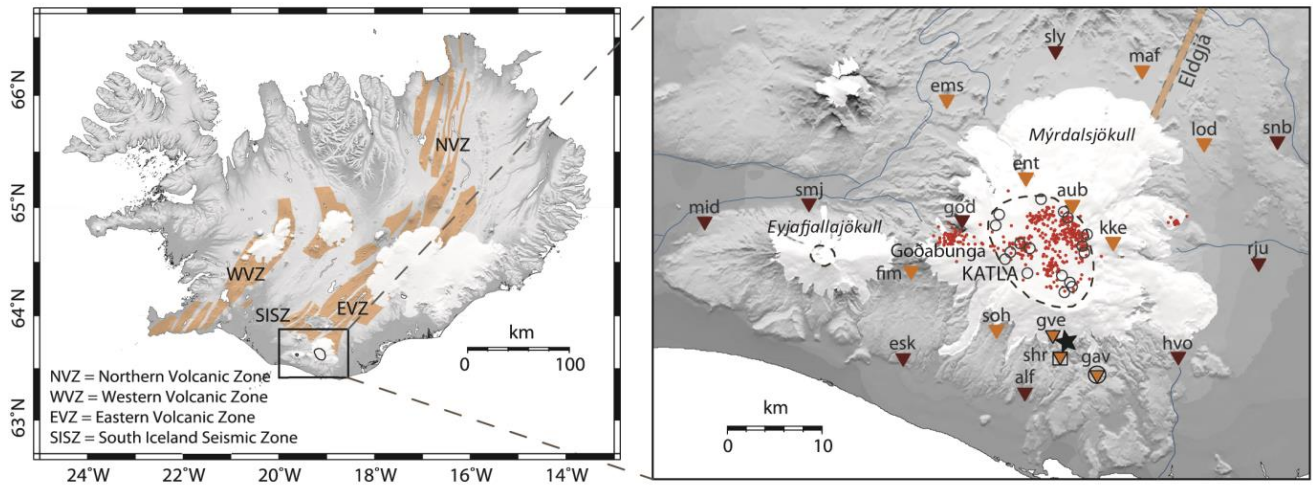

Fig. 1. Map of Iceland showing the different volcanic systems (in orange; from Einarsson and Sæmundsson, 1987) and the different segments of the plate boundary. In the inset, the seismic network and main seismic and geological features of Katla are shown. Dark brown triangles: permanent IMO seismic stations. Orange triangles: temporary Uppsala University seismic stations operating between July 2011-August 2013 (no outline) and between July 2014-August 2014 (black outline). The black open circle marks a temporary station (GAV) that has been operating on both time periods. The black square at station SHR on the south flank indicates the location of a GPS station operating in summer 2014. Red dots: seismic clusters at Katla before July 2011. These are mostly localized in two distinct source areas, within the caldera and on the western flank at Goðabunga. The star marks the new cluster on the south flank. The Katla and Eyjafjallajökull caldera rims are outlined by dashed lines. Open circles correspond to ice cauldrons (Guðmundsson et al., 2007). White areas are glaciers. To the NE, the location of Eldgjá fissure is shown.

## 3. The unrest episode of July 2011

Between August 2010 and July 2011, most of the ice cauldrons on the Mýrdalsjökull glacier uplifted by 6-8 m, interpreted by Guðmundsson et al. (2013) as water accumulation under the glacier. The greatest rise, 11-12 m was observed at cauldron 16 (Fig. 2; Guðmundsson et al., 2013).

In July 2011, the seismicity intensified at Katla and remained high until winter. The unrest culminated on July 8-9th when a burst of tremor was recorded by the Icelandic Meteorological Office (IMO) seismic network. No signs of eruption breaking the ice were seen, but a jökulhlaup drained from Kötlujökull and deepening of some ice cauldrons was observed on the surface of Mýrdalsjökull in the southern and eastern parts of the caldera (Fig. 2).

The tremor burst lasted for about 23 hours, beginning at about 19:00 GMT on July 8th. The jökulhlaup (~18 million m³) swept away the bridge over Múlakvísl river around 05:00 GMT on July 9th, one hour after rising water level was detected at the gauging station Léreftshöfuð, located a few km south of Kötlujökull (IMO, 2011) and ~6 km upstream from the bridge (Fig. 2). Another gauging station, located on the bridge over Múlakvísl river on the main road 1 (Fig. 2), began to show increased conductivity early on July 8th (IMO, 2011).

This unrest episode has been interpreted in association with volcanic processes, such as enhanced geothermal activity, shallow magma intrusion or possibly a minor subglacial eruption (Sgattoni et al., 2015).

The earthquake activity accompanying these events was mainly concentrated inside the caldera, mostly in its south-eastern part. The tremor also originated inside the caldera, in a similar location to the earthquakes (Sgattoni et al., 2015). In addition, a new source of seismic events was activated on the south flank, near Gvendarfell rise, at the southern edge of Mýrdalsjökull glacier (Hafursárjökull), around 4 km south of the southern caldera rim (Fig. 2).

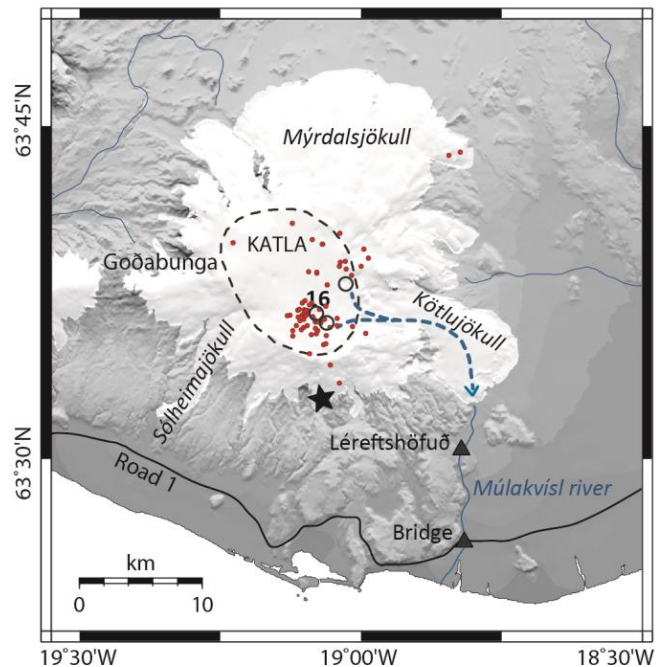

Fig. 2. Map of Katla showing the features related to the July 2011 unrest. Black open circles are the ice cauldrons that deepened during the unrest. Number 16 is the cauldron that showed biggest changes before and during the unrest. The dashed blue arrow shows the presumed flood path. Red dots are the earthquakes that occurred on July 8th and 9th. The 2 gauging stations are marked with black triangles; the southern one corresponds to the bridge over Múlakvísl river. The star marks the southern seismic cluster which is studied herein.

## 4. Seismic network and data

Following the eruption of Eyjafjallajökull volcano in 2010, the IMO augmented the seismic monitoring network around Katla from 5 to 9 stations. Moreover, 9 temporary stations were deployed by Uppsala University between July 2011 and August 2013.

Most of the stations were equipped with broadband sensors, 5 Guralp ESPA , 4 Guralp CMG3-ESPC and 1 Geotech KS-2000(LLC), all with flat response from 60 s to the Nyquist frequency (50 Hz). 5-second Lennartz populated the remaining 8 stations. Data were recorded and digitized with Guralp and Reftek systems at 100 sps. Stations were powered with batteries, wind generators and solar panels. All the instruments recorded in continuous mode, but some technical problems (e.g. power failure) mainly due to harsh weather condition (especially in winter time), prevented some stations from working continuously during the whole operation time.

During the operation time of the dense network described above, the closest station to the Gvendarfell cluster, object of this study, was GAV, around 6 km away. Therefore, the hypocentral location was poorly constrained, especially in depth (uncertainty on the order of a km). Since this seismicity continued with similar characteristics, we decided to install two additional stations, around 1 and 2 km away from the source, during the summer of 2014 to improve the hypocentral location of the cluster. Station GAV was also reactivated to minimize biases in hypocentral estimations caused by the change in network geometry. In addition, a GPS station was deployed during the same period (Fig. 1).

## 5. Waveform characteristics

The Gvendarfell seismicity is characterized by small magnitude (~ -0.5-1.2 $M_L$), long-period earthquakes with an emergent P wave and unclear (usually not identified) S wave (Fig. 3). All events have remarkably similar, nearly identical waveforms with correlation coefficient ≥ 0.9 at the nearest stations. Fig. 4 shows examples of waveforms throughout the whole time period, starting from March 2011, when the first small events were recorded. The similarity of the waveforms is striking and only slight changes can be noticed, from August 2012. The frequency content is narrow banded around 3-4 Hz at the closest station GVE and nearly monochromatic around 3 Hz at most other stations (Figs. 3 and 4).

The signals are characterized by a number of distinct seismic phases that we tried to interpret with particle motion plots. This however didn't help to discern and understand the different phases, which appear to be generated by interference of direct and scattered waves propagating in a heterogeneous medium. The complexity of the waveforms and their duration are therefore highly dependent on the distance from the source. A 6-7 sec long wave-train recorded at ~1 km distance, becomes more than 20 sec long, at 30-40 km. This is indicative of strong path effects, generating a number of secondary phases that increase the complexity of the seismograms (Fig. 5).

Another interesting feature is that the larger events were often preceded by smaller events, only recorded by the closest station GVE. At more distant stations, the small events disappear into the background noise (Fig. 3).

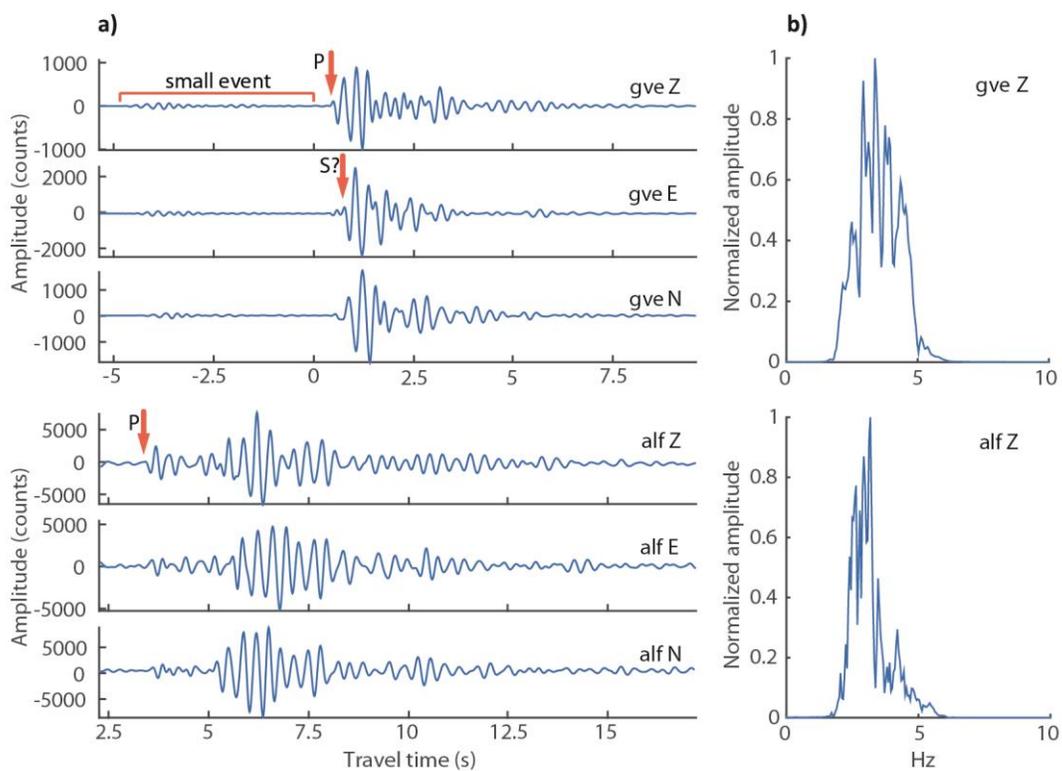

Fig. 3. a) Example seismograms of two different events at stations GVE and ALF. Notice the small event before the main one at GVE. The amplitude unit is digital counts, proportional to velocity. The arrows mark the P and S wave arrivals, where identified. b) Normalised amplitude spectra of the Z component at stations GVE and ALF.

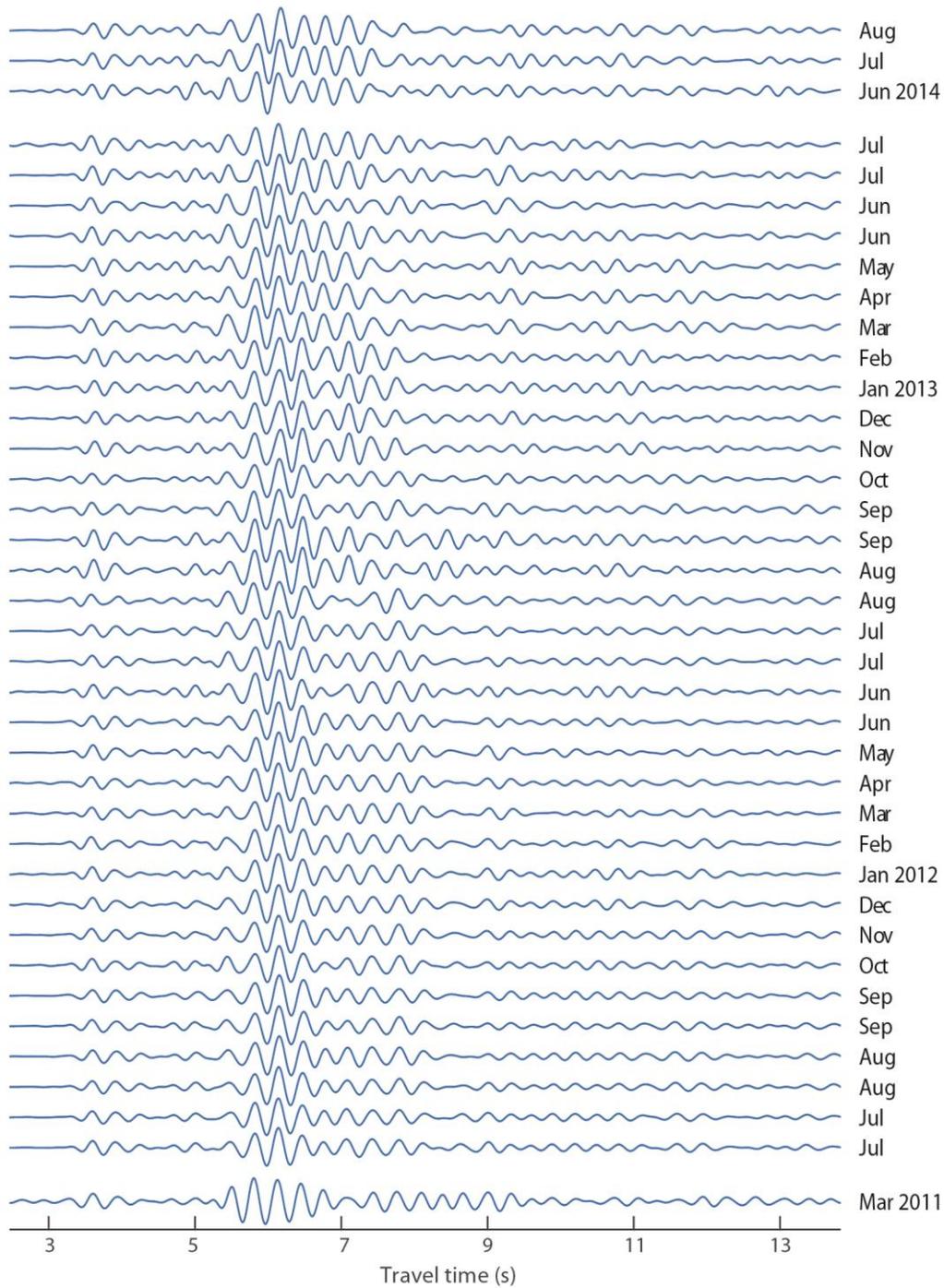

Fig. 4. Example waveforms of the Z component at station ALF (~6 km from the source) throughout the entire period of study, showing their similarity. One event per month is shown for cold seasons and two events per month for warm seasons.

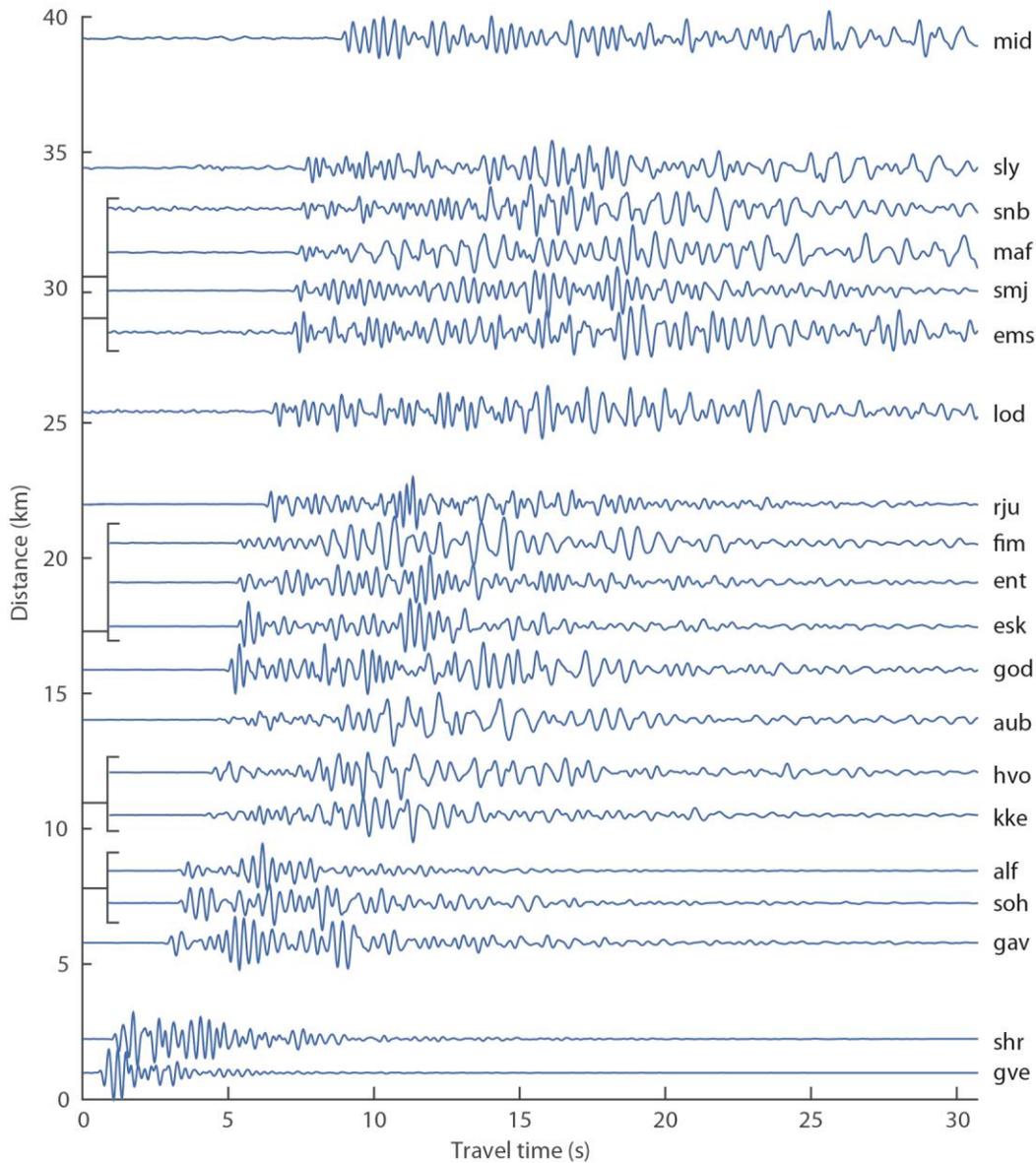

Fig. 5. Stacked Z component seismograms of all events recorded at each station. The approximate epicentral distance is reported on y-axis.

## 6. Time evolution

Visual observation of the seismic data indicated that a significant improvement in detectability could be achieved, compared with the IMO catalogue. As the waveforms are highly repeatable, cross-correlation of a sample waveform with continuous data was applicable to improve the event detection (cf. Lindblom et al., 2015 for a description of the correlation method). For this purpose we used data from station ALF, as it is close to the source and has been working continuously since February 2011. The best event in terms of signal to noise ratio was chosen from the IMO catalogue as a reference event (occurred on Oct 10th, 2011). Its waveform was band-pass filtered between 2-4 Hz and a 3 second window starting 0.5 sec before the P wave arrival was cross-correlated with the continuous data from February 2011 to August 2013. Around 1800 events were detected with a cross-correlation coefficient higher than 0.9 (mostly ≥ 0.95), significantly more compared to the 720 events of the IMO catalogue (Fig. 6). All the events were checked for possible false detections and visual inspection of 3 months of data proved that the cross-correlation analysis detected all visible events successfully. Moreover, in order to check whether other similar events had occurred also before 2011, data from station ESK (operating at Katla before the network was augmented) were used for cross-correlation with continuous data from 2010. In this case, no events were detected that matched a minimum correlation coefficient threshold set as 0.7. This confirms that seismic activity at Gvendarfell was absent before 2011.

The local magnitude ML of the events was evaluated by calibrating the amplitude at ALF with the IMO magnitude estimation (moment magnitude Mw and local magnitude ML). The events overlapping between our improved catalogue and the IMO catalogue were used as a reference. Fig. 7 shows a logarithmic plot of the maximum amplitude at ALF versus ML, showing a clear linear relation. A line fitting of the ML plot was used to translate ALF amplitudes (from our catalogue) into magnitudes (Fig. 7a).

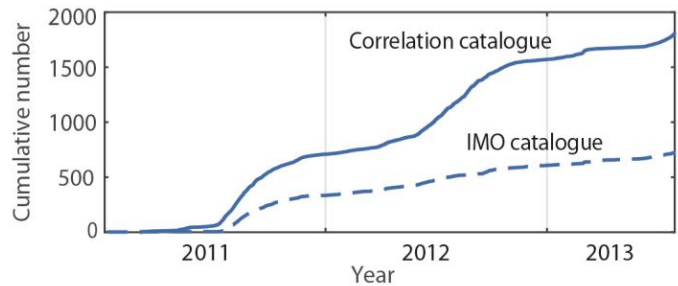

Fig. 6. Cumulative number of events at Gvendarfell. The event detection of the catalogue obtained with cross-correlation is highly improved. A seasonal variation, with more events during warm seasons (corresponding to the steeper segments of the curve), is also clear on the top curve.

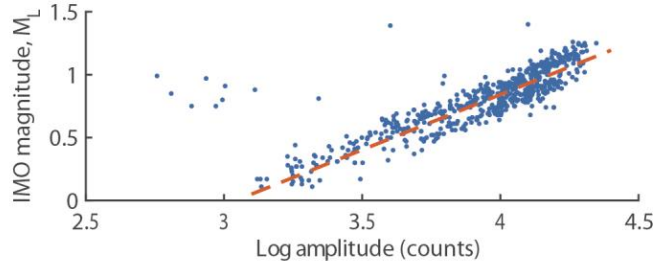

Fig. 7. Magnitude calibration. The maximum log amplitude at station ALF is plotted for all events of our catalogue overlapping with IMO catalogue. Linear correlation between ALF amplitude and IMO local magnitude, $M_L$. The dashed line is the line fitted to the data to convert amplitudes to $M_L$.

The first Gvendarfell events detected at ALF occurred in March 2011 and were very small, around 0.2 ML. A sudden increase in magnitude occurred on July 7th and a striking time feature started on this day: a regular time pattern with 6 events per day at 4 hour intervals began a few hours before the tremor burst of the 2011 unrest episode (Fig. 9). Before that time, this pattern was not observed. A seasonal variation in the event rate is also observed, with maximal activity in late summer 2011, 2012 and 2013, which could not be discerned from the IMO catalogue (Figs. 6, 8a and 9). This correlation is not as clear in the cumulative moment release plot (Fig. 8b), showing a sharp increase in the summer of 2011 and less clear increases in the following summers. The regular event rate is observed also during periods of lower activity, with occurrence frequency gradually decreasing to 1 event every 1-2 days in winter and then increasing again to the maximum rate (6 events per day) during summer (Fig. 9). No diurnal variability or correlation with precipitation rates have been observed. This pattern is seen for the whole time period analyzed in this article (July 2011-August 2014), although the clearest regularity is observed during the summer of 2011. Some exceptions to this general pattern can be noticed, for example in March 2013 (Fig. 8a), when a small swarm of events was recorded, not correlated with any volcanic/geothermal or meteorological event. According to the latest data from IMO, the seismic activity at Gvendarfell seems to be fading out and became insignificant in February 2015.

The amplitude behavior with time also has interesting features. It increased by a factor of 10 at the onset of the tremor and stayed high for a year (Fig.8). After 1 year, the amplitude decreased by a factor of 2. A bimodal size distribution is also observed, with small events occurring in the background, mainly during periods of high seismic activity (Figs. 8a and 9). It is worth noticing that the regular time pattern is mainly observed for the class of larger events, whereas smaller events seem to occur at more random time intervals (Fig. 9).

Fig. 10 shows the relation between the magnitude and the cumulative number of events. The size of the events does not follow the conventional empirical relation of Gutenberg-Richter (logN = a – bM). The bimodal size of the events shows up clearly in the non-cumulative plot in Fig. 10.

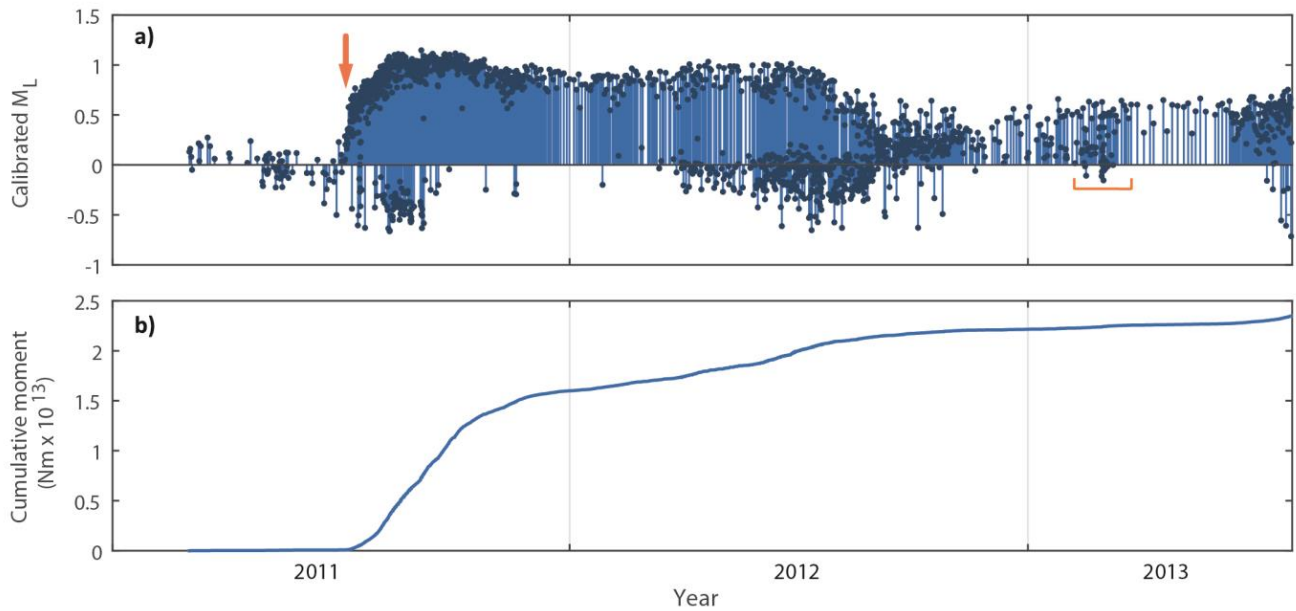

Fig. 8. a) Magnitude ($M_L$) - time evolution of the Gvendarfell seismic sequence between January 2011 and July 2013. The arrow indicates the time of the tremor burst (July 8th-9th). The March 2013 swarm is also outlined (orange squared line). b) Cumulative seismic moment release in the same time interval as panel (a). The moment-magnitude relation used is $logM_0 = 1.5M + 9.1$, where $M_0$ is the moment and M is the magnitude. As we don't have an estimation of Mw, we used $M_L$ in this relation. Therefore, our estimates of seismic moment are uncertain.

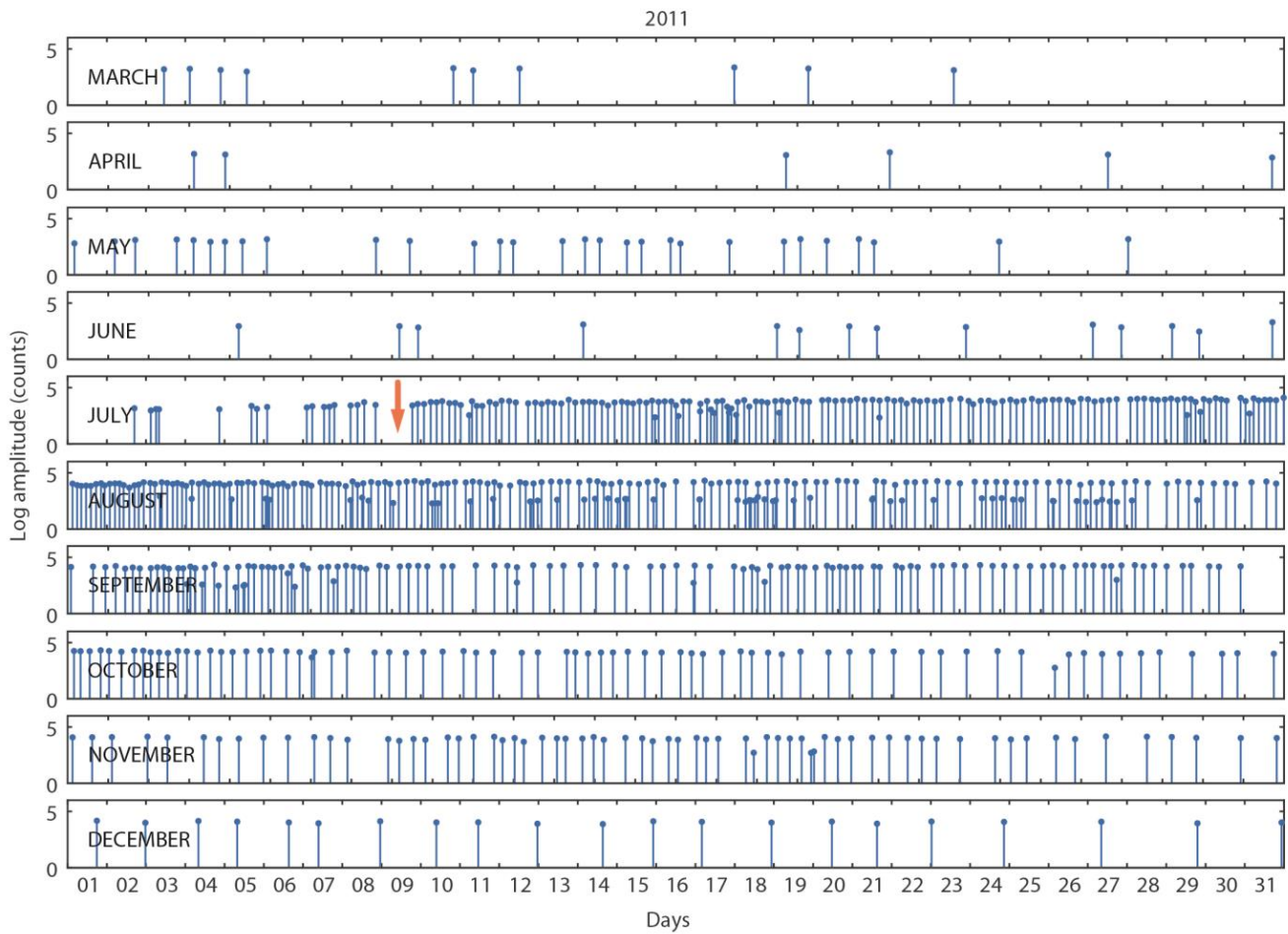

Fig. 9. Time evolution of the Gvendarfell seismic sequence in 2011. The arrow indicates the time of the tremor burst, same as Fig. 8, corresponding to the beginning of the regular time pattern. Inter-event times vary between 4 hours in July-August to 1-2 days in December.

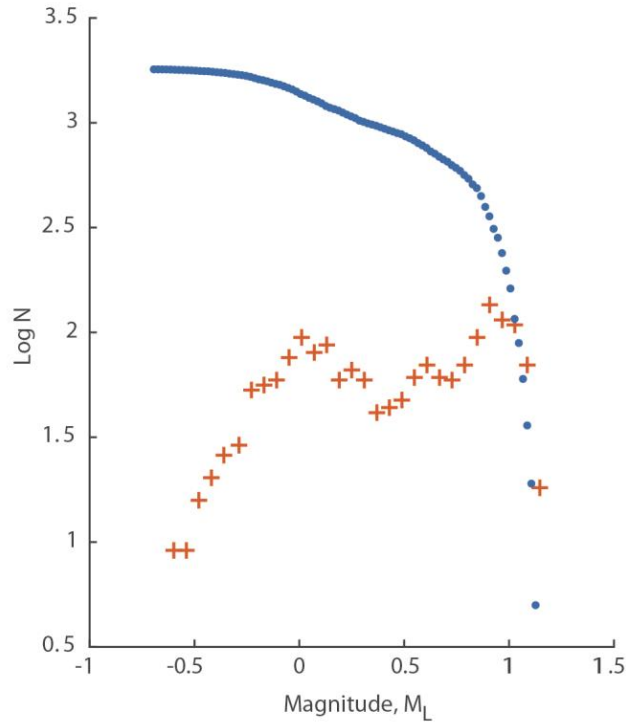

Fig. 10. Magnitude ($M_L$) – number (log N) relation, both cumulative (dots) and non-cumulative (crosses).

## 7. Absolute location

The absolute locations from the IMO catalogue show a cloud of hypocenters, dispersed over a large area, several km in diameter (Fig. 11) around the Hafursárjökull glacial valley, on the southern side of Mýrdalsjökull glacier. However, the similarity of all the waveforms suggests a much smaller distribution of foci. Reported uncertainty is on the order of 1 km in the horizontal and several km in the vertical, i.e. smaller than the distribution of locations, suggesting that significant and variable (because of changes in station geometry) bias due to three-dimensional heterogeneity is present in the locations.

We selected the best recorded events during the summer of 2014 (when 2 additional stations were deployed nearby) to be relocated. For this we used a probabilistic, non-linear method, mapping the likelyhood function for each event around the center of the IMO location cloud with an exhaustive grid search (Lomax et al., 2000). The error distributions of arrival-time data and their predictions are assumed to be Gaussian. The error is not known. A distance weighting is assigned for stations farther than 3 km from the source to simulate increasing uncertainty with increasing distance. This defines a data covariance matrix, less an unknown scaling. This scaling is estimated based on the residual variance at minimum misfit and thus represents the combined error of observation and prediction. The velocity model used is one dimensional and based on tomographic results in the area (Jeddi et al. 2015). Scalar station corrections (estimated based on time residuals) are introduced to absorb effects of lateral heterogeneity with an iterative procedure, starting from no corrections and iterating until converging to stable corrections.

The grid search extended over a 5x5 km$^2$ area around the average IMO location of the events recorded in 2011-2013, down to 5 km depth, with 50 m resolution in the horizontal directions and 100 m in the vertical. As the phase composition of the wave forms is unclear (as explained above), particular care was taken when picking the P and S arrival times. The P wave was carefully identified first in the stacked wave forms and then consistently in the individual events (details of the stacking are explained below). The S wave was only picked at the closest station, GVE. The iterative process of relocation and station-correction estimation converged at the second iteration. The resulting rms correction was 0.17 sec.

Fig. 12 shows the resulting combined probability density of the hypocenter locations. The origin of the coordinate system is the average IMO location (at N63º32.772' and W19º06.588') and depth is referred to the average elevation in the source region (at 800 m.a.s.l.). Our relocations do not differ much from the IMO average, except for a ~500 m shift to the east. The distribution of hypocenters is much smaller than in the IMO catalog locations (Figs. 11 and 12a). The uncertainty of the relocations is around 400 m in the horizontal dimensions and 500-600 m in depth, as inferred from the probability distributions of the locations of single events (Fig. 12b). The cluster appears to be located in the shallow sub surface, between 0.5 and 0.9 km depth. It is clear from comparing Fig. 12a, showing the combined distribution of the cluster, and Fig. 12b, showing the distribution for an individual event, that the cluster distribution is dominated by the uncertainty. Therefore, the cluster may be much smaller in extent than Fig. 12a suggests. The eastward shift of the center of mass of the distribution compared to the average IMO location is marginally resolved. The depth is marginally resolved to differ from zero. A clear correlation (trade off) is persistent between depth and easting. This is controlled by the station geometry, with the nearest station GVE, yielding S picks, a few km to the west-northwest. Some trade off exists between depth and origin time, but inclusion of S picks from station GVE reduces that significantly.

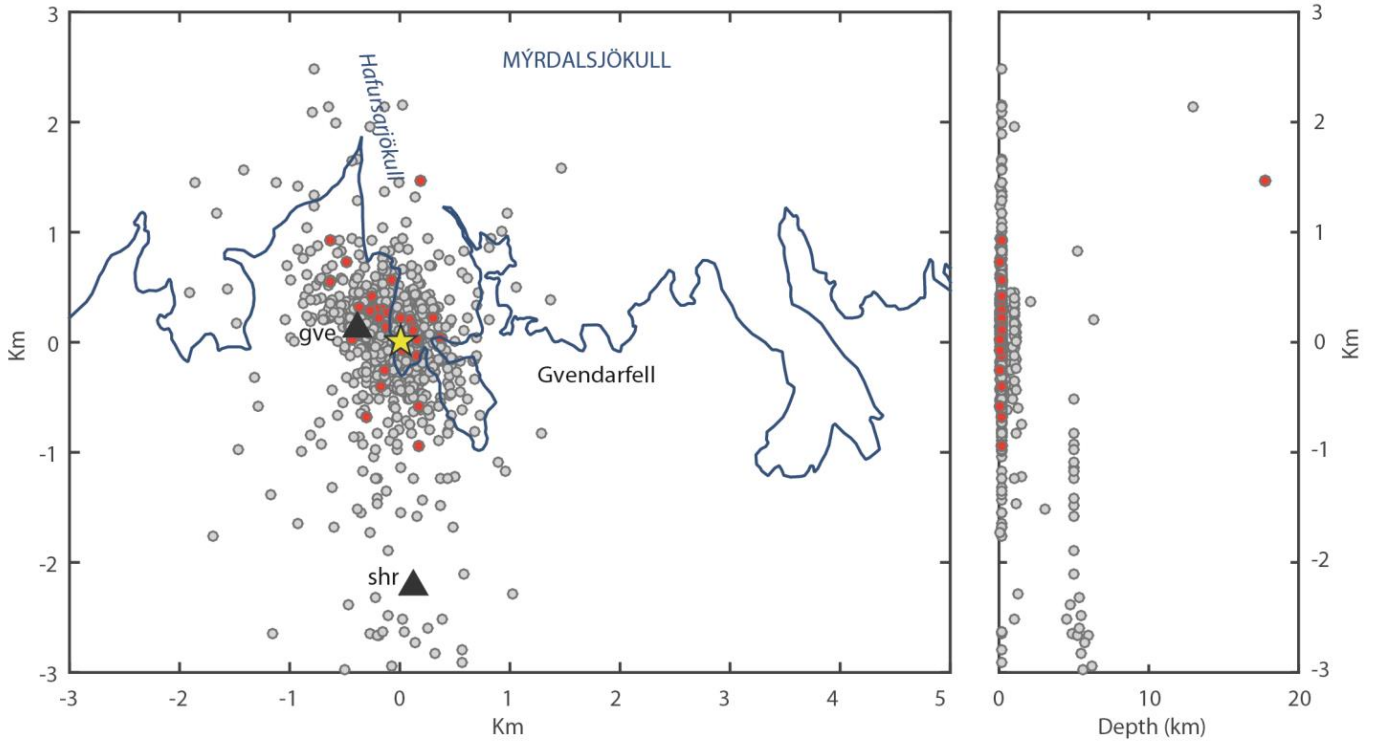

Fig. 11. Locations and depth distribution of the Gvendarfell seismic events, from IMO catalogue: grey = all locations 2011-2013; red = 30 events in summer 2014, same relocated with nonlinear method (see Fig. 12). The star is the average IMO location (N63°32.772', W19°06.588') and corresponds to the origin of the axes of Fig. 12. Black triangles are the seismic stations. The glacier outline (in blue) is derived from LiDAR DEM obtained in 2010 (Jóhannesson et al., 2013).

## 8. Focal mechanisms

Although we do not know what kind of source generates the Gvendarfell events, we tried to obtain some description of it using regular earthquake source analysis (e.g. focal mechanism based on first motion polarities). If the source is different from that of a tectonic earthquake (e.g. geothermal/volcanic processes involving fluids), a shearing type focal mechanism would be inappropriate, but the first motion descriptions still valuable.

The waveforms of the Gvendarfell seismic events are strongly affected by path effects. In addition, the geometry of the network was unfavourable for source studies, with most stations located several km away from the source, mainly because of the glacier. Therefore, it was not considered promising to use a full waveform inversion to determine the source mechanism. Such a method is also very sensitive to erroneous velocity models, particularly at shallow depth (Bean et al., 2008), which strongly affects stations farther from the source. Therefore, an attempt to retrieve the focal mechanism was done based on the first motion polarities. The emergent P onset is difficult to identify for low-frequency events. Therefore, the signal-to-noise ratio was improved by aligning and stacking all the waveforms at each station (20 stations in total). This was possible thanks to the extreme similarity of the waveforms. Each waveform was weighted according to its signal-to-noise ratio and the uncertainty of the stack evaluated as variance of the weighted mean. The first arrivals, however, remained unclear at many stations. A closer insight in the zoomed P waves, showed that the first wiggles of all waveforms are highly correlated between different stations. Therefore, by taking the clearest onset (positive polarity at the closest station GVE) as reference, we were able to interpret all the others. We did this by correlating the reference GVE onset with the stacked P onset at all other stations, first the original ones and then the reversed.

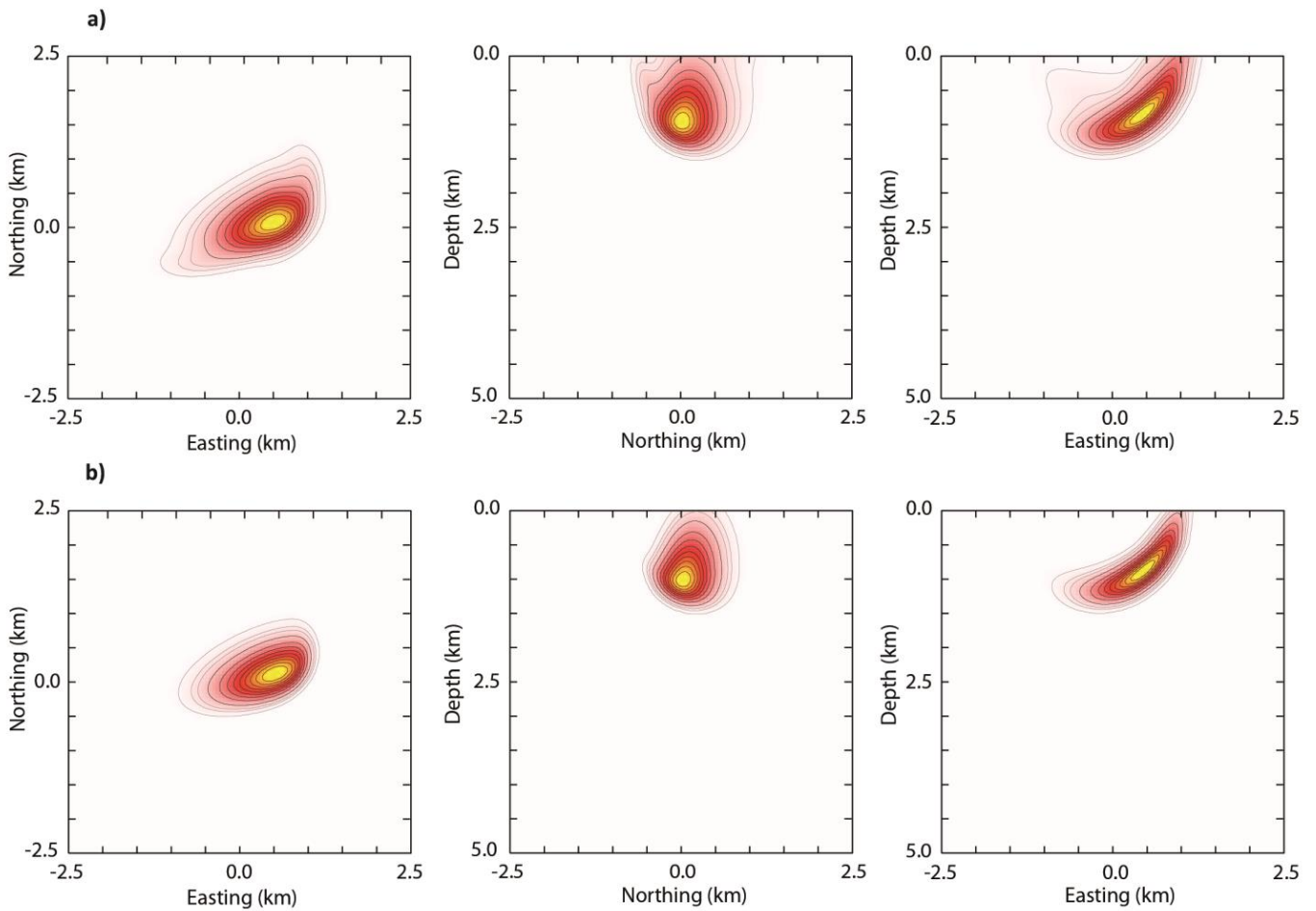

Fig. 12. a) Non-linear locations of the Gvendarfell seismic events: combined probability density of all locations (34 events in total occurred in summer 2014). b) Probability density of the location of one example event, to show the single event uncertainty. The scale is normalised and the density ranges from 1 (yellow, in the centre of the distribution) to zero (white). The black contour lines correspond to 0.9 – 0.1 with 0.1 spacing. The grey contour lines are 0.05, 0.03, 0.01.

The stations resulting in the best correlation for the original onset were assigned a positive polarity and vice-versa. The reason for this is intuitive and looks clear from Fig. 13, showing the resulting polarities and the beginning of the seismogram for all stations.

The interpretation of the first motions, however, is not trivial and no unique solution of the focal mechanism can be obtained. This arises mainly from the poor coverage of the focal sphere. Also, as the source is located in the shallow, low-velocity layers, take-off angles can vary considerably and strong scattering effects due to the heterogeneous medium are expected to affect azimuth angles. A tentative fault plane solution is drawn, consistent with normal faulting. This is however not well constrained and other mechanisms, with different combinations of CLVD (Compensated Linear Vector Dipole) and volumetric components might be invoked. Certainly, the polarities are not consistent neither with thrust faulting, nor with pure implosion/explosion.

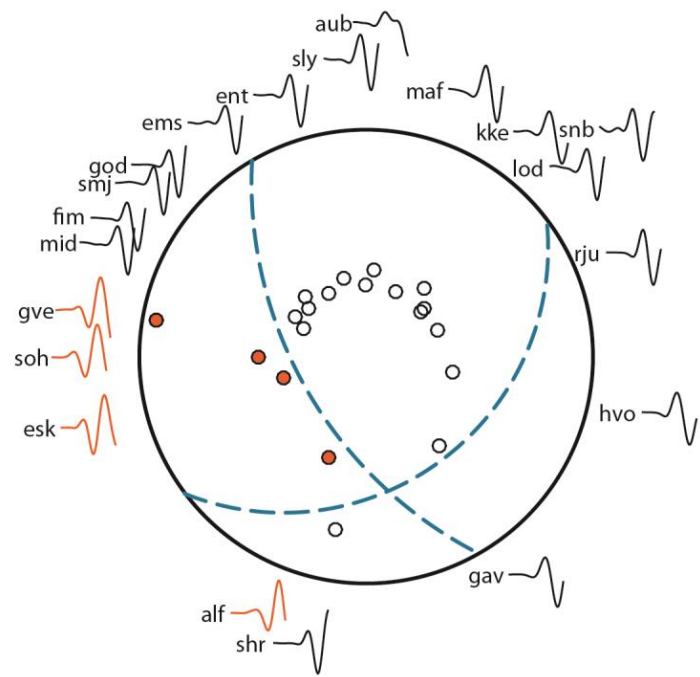

Fig 13. First motion polarities of the Gvendarfell events, for all seismic stations. Orange dots for positive polarities and white for negative. The beginning of the Z component seismogram for each station is also plotted, with colour corresponding to the polarity. A tentative fault plane solution consistent with normal faulting is drawn.

## 9. Discussion

The Gvendarfell seismic events have low frequency content, observed at all stations, from the closest (around 1 km) to the farthest (around 40 km). Path effects seem to be important at most stations, as seen from increasing complexity of the waveforms with distance from the source. Moreover, the presence of the glacier might also play a role as a filter for high frequencies (Weaver and Malone, 1979; Métaxian et al., 2003). If this was dominant, we would expect the same effect for the caldera events as well, where the glacier is much thicker, up to 700 m, compared to few tens of meters at Hafursárjökull above the Gvendarfell seismic source. However, we point out that other shallow events originating inside the Katla caldera contain a broader spectrum of frequencies, up to 10-15 Hz, with higher frequencies not completely attenuated by path effects. We therefore interpret the low frequency content of the Gvendarfell events as a source property and classify them as LP events. Moreover, other features of the waveforms are similar to those of LP events at other volcanoes: i) similarity of waveforms, ii) emergent onset of P wave, iii) unclear S wave, iv) narrow peaked spectra with typical frequencies between 0.5 and 5 Hz (Chouet, 2003).

As the depth of the cluster is not distinguishable with certainty from the surface and the location corresponds to a glacier stream, glacial processes must be considered also as possible sources of the Gvendarfell seismicity. Glacial earthquakes can be associated with glacier sliding, e.g. stick-slip ice motion (Weaver & Malone, 1979; Ekström et al., 2006; Wiens et al, 2008; Thelen et al., 2013) or with ice avalanches and ice falls (Weaver & Malone, 1979; Caplan-Auerbach and Huggel C., 2007; Roux et al., 2008; Jónsdóttir et al., 2009). Resonance of ice cracks (Métaxian et al., 2003) or resonance of the glacier (Wolf and Davies, 1986) can also generate LP events, but for the reasons discussed below, we consider resonance processes unlikely. A glacial origin of these events could be in accordance with the seasonal variation observed, as it is usual for glacier seismicity to peak during warm seasons (e.g. Wolf and Davies, 1986; West et al., 2010; Moore et al., 2013). Glacier movements are enhanced due to summer melting, which in Iceland peaks in July and August, after the summer thaw. An example of regular and repetitive glacial seismicity over years has been observed in a Transantartic Mountains glacier (Winberry et al., 2015). The size (likely a few hundred meters) and slip-rate (50 m per year) of that glacier are however much bigger compared to Hafursárjökull glacier. Hafursárjökull is a small, stagnant alpine glacier, not more than few tens of meters thick, and is unlikely to be capable of producing such a persistent seismicity. Fig. 14 shows a picture of the glacier, highlighting its small size and showing its actual margin, compared to the outline drawn in Fig. 11. Alpine glaciers' seismicity is usually highly variable in amplitude, recurrence interval, waveforms and often shows changes on diurnal basis and correlation with precipitation rates (Thelen et al., 2013). None of these characteristics are seen at Gvendarfell, where only one family of nearly identical seismic events has been observed for more than 3 years. In addition, if a relation existed between the Gvendarfell seismicity and glacial processes, it would be difficult to understand why similar events are not observed at more active glacier streams nearby or at other glaciers in Iceland. As a final consideration, the time association with the unrest episode in July 2011 seems easier to reconcile with volcanic or geothermal processes.

Due to their association with eruptions, it is commonly thought that LP events are caused by fluid movements within the volcano, for example magma injection into fractures or crypto-dome formation (Okada et al., 1981; Cruz and Chouet, 1997; Rowe et al. 2004; Neuberg et al., 2006), or processes related to hydrothermal systems (Moran et al., 2000; Saccorotti et al., 2007; Guðmundsson and Brandsdóttir, 2010; Jousset et al., 2010; Matoza and Chouet, 2010; Arciniega-Ceballos et al., 2012).

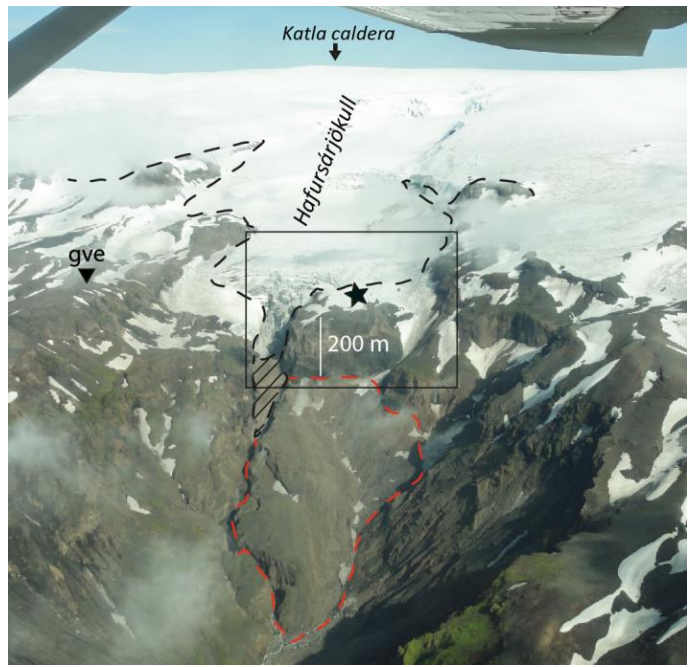

Fig. 14. Aerial view of Hafursárjökull glacier. The red dashed line is the glacier outline shown in Fig. 11. The black dashed line is the actual glacier limit as it is at the time we are writing (August 2015). The area marked with black lines is ice covered by debris. The star corresponds to the approximate location of the center of the absolute hypocenter locations and the box crudely indicates its uncertainty. The approximate location of the seismic station GVE is also drawn with a black triangle.

The narrow-banded frequency range of LP seismograms at volcanoes is often associated with resonance processes in which the dominant frequency of the source is related to specific geometries and/or fluid characteristics. In particular, monochromatic, volcanic LP events may be interpreted in terms of resonance of fluid-filled cracks (e.g. Chouet, 1996; Kumagai and Chouet, 1999; Saccorotti et al., 2007). However, a resonance process is expected to generate a long coda of decaying harmonic oscillations, which is not observed in the Gvendarfell events. While most stations record a long-duration signal, resembling resonance, the proximal observation from GVE station is short in duration and indicative of a source more pulse-like rather than resonating. This feature of LP events has been described by Bean et al. (2014) and explained as a strong path effect, very pronounced for shallow sources in the low-velocity, near-surface layers. They carried out a dynamic rupture simulation and showed that long-period signals can be generated by slow-rupture failure in unconsolidated volcanic materials. This indicates that LP events are not necessarily generated in association with fluid movements. Their simulations also show a spread of hypocenters, which seems to be not the case of the Gvendarfell seismicity, generated at a very small source, stable with time and always producing the same waveform. At Gvendarfell there is neither variability of waveforms, nor evidence for multiple structures acting as seismic sources, as described by Bean et al. (2014).

The high repeatability of waveforms and regular occurrence of LP events characteristic of the Gvendarfell seismicity have been observed at other volcanoes worldwide (Cruz and Chouet, 1997; Ramos et al., 1999; Green and Neuberg, 2006; Rowe et al. 2013), often associated with rising domes (e.g. Okada et al., 1981; Neuberg et al., 2006) or with the interaction of hydrothermal and magmatic systems. For example, repetitive LP events at Popocatépetl were associated with the non-destructive process of repetitive injection of hydrothermal fluid into a fracture resulting in a sudden discharge when a critical pressure is reached (Arciniega-Ceballos et al., 2012). Matoza and Chouet (2010) interpreted repetitive LP events occurring at Mt. St. Helens with regular inter-event time spacing in terms of rapid heating of water and dissolved volatiles in a shallow hydrothermal crack, with events triggered by phase changes from liquid to vapor. In all cases, LP events have been observed either prior to or during eruptions, their repeatability lasts for days to months and the regular time intervals between successive events are on seconds to minutes scale. The Gvendarfell events, instead, did not precede or accompany an eruption (at least not in the same area), although they started in coincidence with a tremor event recorded inside the caldera. Moreover, the same waveforms have been recorded for around 3.5 years, occurring with regular inter-event times ranging from hours to days, depending on the season. These features make the Gvendarfell seismicity a unique case.

The Gvendarfell events are also different compared to the other LP events recorded at Katla, most of which are located at Goðabunga, on the western flank of the volcano. The other Katla LP events have a wider spectrum of frequencies, with higher frequencies in the beginning of the waveform decreasing towards the end, as opposed to the nearly monochromatic waveforms at Gvendarfell. In addition, no other cluster at Katla is composed of only one family of waveforms and no other cluster has a similar regular time pattern. The seasonal correlation also differs: the Goðabunga activity peaks later in autumn (Einarsson and Brandsdóttir, 2000; Jónsdóttir et al., 2009) as opposed to July/August at Gvendarfell and the caldera events have a less pronounced seasonality.

The bimodal size distribution is also peculiar and is reflected in the magnitude-cumulative number curve, where it is difficult to identify a unique b-value (Fig. 10). Small earthquakes clearly dominate, implying that the source size is restricted and not capable of producing events larger than a threshold size. In addition, the two classes of events occur with different time patterns, which indicate two separate phenomena, although the waveforms are highly correlated. A similar case, with minor events occurring in the background of larger LP events, was observed at Mt. St. Helens during the 2004-2008 eruption period and interpreted in terms of less energetic fluid to gas phase changes, involving a smaller volume of fluid in a shallow hydrothermal crack (Matoza and Chouet, 2010). Non-linear magnitude-frequency relationship is commonly observed at volcanoes, e.g. the earthquakes accompanying the dyke injection at Krafla in September 1977 (Brandsdóttir and Einarsson, 1979).

Since we couldn't model the source mechanism, for the reasons discussed in section 8, we are not able to discuss in detail the physical process generating the Gvendarfell events. However, the striking time evolution, together with the repeatability of the waveforms over years, are indicative of a remarkably stable, non-destructive process repeating itself at regular time intervals, modulated by seasons, for an unusually long time. Although glacial processes cannot be completely ruled out, we regard volcano-related processes as more likely to generate such a stable seismicity.

Among volcanic processes, both magma-related (e.g. magma intrusion, dome formation) and non-magma related hydrothermal processes (e.g. fluid instabilities, thermal cracking) are candidate sources. The former interpretation fits with the geological evidence of silicic extrusive bodies at the caldera rim and on the eastern and southern flanks of Katla (Lacasse et al., 2007; Jóhannesson and Sæmundsson, 2009). This makes a rising shallow viscous magma body a plausible geological processes in the Gvendarfell area. A normal faulting mechanism with a dilatational component, consistent with the first motion polarities shown in Fig. 14, could fit with the hypothesis of an intruding dike. The time association with the tremor episode of the 2011 unrest would also fit with such magma-related interpretation. However, the GPS station operating during the summer of 2014 near the Gvendarfell area did not report evidence of ground deformation to support this hypothesis of a slowly intruding dyke. Moreover, the seasonal correlation of the seismicity appears difficult to reconcile with magma-related processes. Thus, we regard hydrothermal processes as more likely than magma movements to explain this seismicity. A minor hydrothermal system, lasting for about 3.5 years, might have been activated on Katla's south flank during the unrest episode in July 2011. The regular time pattern of the seismicity could be associated with a steady process of cyclic heating and cooling of the fluid phase in a hydrothermal crack system with LP events generated by phase changes between liquid and vapor (Matoza and Chouet, 2010). In this scenario, the seasonal correlation of the seismicity could be explained in terms of varying supply of ice melt water to the hydrothermal system, with peaks correlating with the warm seasons. Less energetic phase changes, involving smaller volumes of fluids, might explain the subset of smaller events. The first motion polarities, although not univocally interpretable, can be consistent with non-double-couple mechanisms observed at other geothermal areas such as Hengill in Iceland (e.g. Miller et al., 1998). However, we point out that there is no visible evidence of geothermal activity, new or old, in the area. The new hydrothermal system, therefore, has to be entirely concealed. Furthermore, a new hydrothermal system needs a new heat source. A small dyke injection into the southern caldera wall or changes to a permeable crack system in conjunction with the thermal event in July 2011 are possible scenarios. A summary of possible interpretations and corresponding pros and cons is reported in Table 1.

| Process | Pros | Cons |
|---|---|---|
| **GLACIAL** <br> e.g. glacier sliding, ice-falls | - seasonal correlation <br> - shallow depth | - small, stagnant glacier <br> - no correlation with precipitation rates <br> - association with unrest episode <br> - stability of the process over long time <br> - depth (?) |
| **VOLCANIC** <br> (magma involved) <br> e.g. dome rising, viscous magma injection | - geological evidences: silicic extrusive bodies at the caldera rim and south/east flanks <br> - association with unrest episode <br> - normal faulting mechanism (?) | - seasonal correlation <br> - no ground deformation detected |
| **HYDROTHERMAL** <br> (no magma involved) <br> e.g. phase changes of the geothermal fluid, thermal cracking | - shallow depth <br> - high repeatability <br> - association with unrest episode <br> - seasonal correlation can be explained <br> - regular time pattern can be explained | - no evidence of geothermal activity |

Table 1. Summary of pros and cons of the three suggested interpretations of the source process of the Gvendarfell seismic events.

## 10. Conclusions

Since July 2011 the seismicity pattern at Katla volcano has shown changes: a new cluster of shallow, repeating LP seismic events has been observed on Katla's south flank, at the southern edge of Mýrdalsjökull glacier, 4 km south of the caldera rim. The onset of this seismicity coincided with an unrest episode culminated in a glacial flood from the south-east rim of the glacier, on July 9th, 2011. The seismicity on the south flank had never been observed before and continued for around 3.5 years with the same features.

Since these seismic events are located in a glaciated area, both volcanic and glacial processes must be taken into account as possible sources. Because of the characteristics of the small glacier stream and because of the remarkable stability of the main features of the seismicity over years, we regard volcano-related processes as more likely to generate this seismicity. However, we cannot rule out a glacial source and this study highlights the difficulty and the importance of discriminating glacial and volcanic sources at subglacial volcanoes.

Although they share some common features with LP earthquakes at other volcanoes worldwide, the seismic events we have described represent a peculiar case study because of their temporal behavior and because they did not accompany an eruption. We have not found a similar case in the literature. They also differ significantly from other Katla's LP events.

Among volcano-related process, we suggest a shallow hydrothermal system is more likely than magma movement to explain this seismicity, mainly because the clear seasonal correlation is easier to reconcile with a process involving water. The extremely regular time pattern over a long time (at least 3.5 years) together with the similarity of all the waveforms, point to a stable, non-destructive source mechanism over time. We regard the regular seismic rate, modulated by seasonality, as the most striking feature of this seismicity. Therefore, we look for a steady source process, in both location and mechanism, in which some critical parameter induces the regular time interval between events. This might be related to a steady process of heating and cooling of a fluid phase, in a geyser-like process. As the fluid phase is supplied by the glacier, a seasonal correlation can be expected as a response to the summer ice melting. Our hypothesis is therefore that a small, shallow hydrothermal system might have been activated on Katla's south flank, in coincidence with the 2011 unrest episode. As a power source, we suggest either a short lived dike intrusion towards the south flank or a crack connection to a heat source established during the unrest. However, no evidence of old or new hydrothermal activity has been seen in the area.

Further studies, such as relative relocation of the hypocenters and further insights into the tremor episode of July 2011, will help interpreting the source processes and volcanological implications of this peculiar cluster of LP events on the south flank of Katla.

## Acknowledgements

The authors would like to thank the Icelandic Met Office for access to waveform data and catalogue data of the Gvendarfell events. The temporary deployments producing data for this study were supported by CNDS (Centre for Natural Disaster Science, www.cnds.se) at Uppsala University and the Volcano Anatomy project, financed by the Icelandic Science Foundation. This work was funded by the University of Bologna, University of Iceland and Uppsala University, as a part of a joint PhD project. We thank also the sheriff in Vík for logistic support and Vincent Drouin for the flight over the area.

## References

Arciniega-Ceballos, A., Dawson, P., Chouet, B. a., 2012. Long period seismic source characterization at Popocatépetl volcano, Mexico. Geophys. Res. Lett. 39, 1–5. doi:10.1029/2012GL053494

Bean, C., Lokmer, I., O'Brien, G., 2008. Influence of near-surface volcanic structure on long-period seismic signals and on moment tensor inversions: Simulated examples from Mount Etna. J. Geophys. Res., 113, B08308. doi:10.1029/2007JB005468

Bean, C.J., De Barros, L., Lokmer, I., Metaxian, J.P., O'Brien, G., Murphy, S., 2014. Long-period seismicity in the shallow volcanic edifice formed from slow-rupture earthquakes. Nature Geoscience, 7 (1), 71-75.

Björnsson, H., Pálsson, F., Guðmundsson, M.T., 2000. Surface and bedrock topography of the Mýrdalsjökull ice cap, Iceland: The Katla caldera, eruption sites and routes of jökulhlaups. Jökull 49, 29-46.

Bransdóttir, B., Einarsson, P., 1979. Seismic activity associated with the September 1977 deflation of the Krafla central volcano in northeastern Iceland. J. Volcanol. Geotherm. Res. 6 (3), 197-212.

Budd, D.A., Troll., V.R., Dahren, B., Burchardt, S., 2014. Persistent shallow magma storage beneath Katla Volcano. Paper presented at: Goldschmidt Annual Meeting, Sacramento, USA.

Chouet, B.A., 1996. Long-period volcano seismicity: its source and use in eruption forecasting. Nature, 380, 309-316.

Chouet. B.A., 2003. Volcano Seismology. Pure appl. geophys. 160, 739-788.

Cruz, F.G., Chouet, B.A., 1997. Long-period events, the most characteristic seismicity accompanying the emplacement and extrusion of a lava dome in Galeras Volcano, Colombia, in 1991. J. Volcanol. Geotherm. Res., 77, 121-158.

Ekström, G., Nettles, M., Tsai, V. C., 2006. Seasonality and increasing frequency of Greenland glacial earthquakes. Science, 311, 1756– 1758.

Einarsson, P., 1991. Earthquakes and present-day tectonism in Iceland. Tectonophysics 189, 261–279.

Einarsson, P., Brandsdóttir, B., 2000. Earthquakes in the Mýrdalsjökull area, Iceland, 1978–1985: Seasonal correlation and relation to volcanoes. Jökull 49, 59–73.

Einarsson, P., Sæmundsson, K., 1987. Earthquake epicenters 1982-1985 and volcanic systems in Iceland. In Þ.I.Sigfússon, ed. *Í hlutarins eðli*, Festschrift for Þorbjörn Sigurgeirsson. Menningarsjóður, Reykjavík (map).

Green, D.N., Neuberg, J., 2006. Waveform classification of volcanic low-frequency earthquake swarms and its implication at Soufrière Hills Volcano, Montserrat. J. Volcanol. Geotherm. Res. 153, 51–63. doi:10.1007/1-4020-4287-6_4

Guðmundsson, Ó., Brandsdóttir, B., 2010. Geothermal noise at Ölkelduháls, SW Iceland. Jökull 60, 89-102.

Guðmundsson, Ó., Brandsdóttir, B., Menke, W., Sigvaldason, G.E., 1994. The crustal magma chamber of the Katla volcano in South Iceland revealed by 2-D seismic undershooting. Geophys. J. Int. 119, 277–296.

Guðmundsson, M.T., Högnadóttir, Þ., Kristinsson, A.B., Guðbjörnsson, S., 2007. Geothermal activity in the subglacial Katla caldera, Iceland, 1999–2005, studied with radar altimetry. Annals of Glaciology 45, 66–72.

Guðmundsson, M.T., Larsen, G., Sigmarsson, O., 2013. In: Sólnes, J., Sigmundsson, F., Bessason, B. (Eds.), Náttúruvá Á Íslándi. Eldgos og Jarðskjálftar. Viðlagatrygging Íslands/Háskólaútgáfan, p. 2013.

IMO, Icelandic Meteorological Office, 2011. http://en.vedur.is/.

Jakobsdóttir, S.S., 2008. Seismicity in Iceland: 1994–2007. Jökull 58, 75–100.

Jeddi, Z., Tryggvason, A., Guðmundsson, Ó., IMO-SIL monitoring group, 2015. 3D Velocity structure of the Katla volcano - Southern Iceland. Poster session presented at: 26th IUGG General Assembly 2015. Prague, Czech Republic.

Jóhannesson, T., Björnsson, H., Magnússon, E., Guðmundsson, S., Pálsson, F., Sigurðsson, O., Thorsteinsson, T., Berthier, E., 2013. Ice-volume changes, bias estimation of mass-balance measurements and changes in subglacial lakes derived by lidar mapping of the surface Icelandic glaciers, Ann. Glaciol., 54(63), 63–74. doi:10.3189/2013AoG63A422

Jóhannesson, H., Sæmundsson, K., 2009. Geological Map of Iceland. 1:600.000. Bedrock Geology, 1st edition. Icelandic Institute of Natural History, Reykjavík.

Jónsdóttir, K., Roberts, R., Phjola, V., Lund, B., Shomlai, Z. H., Tryggvason, A., Bodvarsson, R., 2009. Glacial long period seismic events at Katla volcano, Iceland. Geophys. Res. Lett. 36, L11402.

Jónsdóttir, K., Tryggvason, A., Roberts, R., Lund, B., Soosalu, H., Bodvarsson, R., 2007. Habits of a glacier covered volcano: seismicity and a structure study of the Katla volcano, South Iceland. Annals of Glaciology 45, 169–177.

Jónsson, G., Kristjánsson, L., 2000. Aeromagnetic measurements over Mýrdalsjökull and vicinity. Jökull 49, 47–58.

Jousset, P., Haberland, C., Bauer, K., Arnason, K., Weber, M., Fabriol, H., 2010. Seismic Tomography and Long-Period Earthquakes Observation and Modelling at the Hengill Geothermal Volcanic Complex, Iceland. Proc. World Geotherm. Congr. 2010 25–29.

Kumagai, H., Chouet, B.A., 1999. The complex frequencies of longperiod seismic events as probes of fluid composition beneath volcanoes. Geophys. J. Int. 138, F7–F12.

Lacasse, C., Sigurðsson, H., Carey, S.N., Jóhannesson, H., Thomas, L.E., Rogers, N.W., 2007. Bimodal volcanism at the Katla subglacial caldera, Iceland: Insight into the geochemistry and petrogenesis of rhyolitic magmas. Bulletin of Volcanology 69, 373–399.

Larsen, G., 2000. Holocene eruptions within the Katla volcanic system, south Iceland: characteristics and environmental impact. Jökull 49, 1–28.

Lindblom, E., Lund, B., Tryggvason, A., Uski, M., Bödvarsson, R., Juhlin, C., Roberts, R., 2015. Microearthquakes illuminate the deep structure of the endglacial Pärvie fault, northern Sweden. Geophys. J. Int., 201, 1704–1716.

Lomax, A., Virieux, J., Volant, P., Berge, C., 2000. Probabilistic earthquake location in 3D and layered models: Introduction of a Metropolis-Gibbs method and comparison with linear locations. In: Thurber, C.H., Rabinowitz, N. (eds.), Advances in Seismic Event Location, Kluwer, Amsterdam, 101-134.

Matoza, R. S., Chouet, B. A., 2010. Subevents of long-period seismicity: Implications for hydrothermal dynamics during the 2004–2008 eruption of Mount St. Helens. J. Geophys. Res., 115, B12206. doi:10.1029/2010JB007839

McNutt, S.R., 2005. Volcanic Seismology, Ann. Rev. Earth Planet. Sci. 33. doi: 10.1146/annurev.earth.33.092203.122459

Métaxian, J.-P., Araujo, S., Mora, M., Lesage P., 2003. Seismicity related to the glacier of Cotopaxi Volcano, Ecuador. Geophys. Res. Lett. 30(9), 1483. doi: 10.1029/2002GL016773

Miller, A.D., Foulger, G.R., Julian, B.R., 1998. Non double-couple earthquakes 2. Observations. Rev. Geophys. 36, 551-568.

Moore, P. L., Winberry, J. P., Iverson, N. R., Christianson, K. A., Anandakrishnan, S., Jackson, M., Mathison, M. E., Cohen., D., 2013. Glacier slip and seismicity induced by surface melt. Geology, 41(12), 1247-1250.

Moran, S.C., Zimbelman, D.R., Malone, S.D., 2000. A model for the magmatic–hydrothermal system at Mount Rainier, Washington, from seismic and geochemical observations. Bull. Volcanol. 61, 425–436. doi:10.1007/PL00008909

Neuberg, J. W., Tuffen, H., Collier, L., Green, D., Powell, T., Dingwell, D., 2006. The trigger mechanism of low-frequency earthquakes on Montserrat. J. Volcanol. Geotherm. Res. 153, 37-50.

Okada, H., Watanabe, H., Yamashita, H., Yokoyama, I., 1981.Seismological significance of the 1977–1978 eruptions and the magma intrusion process of Usu volcano, Hokkaido. J. Volcanol. Geotherm. Res. 9, 311 –334.

Óladóttir, B.A., Sigmarsson, O., Larsen, G., Thordarson, T., 2008. Katla volcano, Iceland: Magma composition, dynamics and eruption frequency as recorded by Holocene tephra layers. Bull. Volcanol. 70, 475–493. doi:10.1007/s00445-007-0150-5

Ramos, E., Hamburger, M.W., Pavlis, G.L., Laguerta, E.P., 1999. The low-frequency earthquake swarms at Mount Pinatubo, Philippines: implication for magma dynamics. J. Volcanol. Geotherm. Res. 92, 295– 320.

Roberts, M. J., Tweed, F. S., Russell, A. J., Knudsen, Ó., Harris, T. D., 2003. Hydrological and geomorphic effects of temporary ice-dammed lake formation during jökulhlaups. Earth Surf. Processes Landforms 28, 723– 737.

Roux, P. F., Marsan, D., Métaxian, J. P., O'Brien, G., Moreau, L., 2008. Microseismic activity within a serac zone in an alpine glacier. J. Glaciol., 54, 157– 168.

Rowe, C.A., Thurber, C.H., White, R.A., 2004. Dome growth behavior at Soufriere Hills Volcano, Montserrat, revealed by relocation of volcanic event swarms, 1995-1996. J. Volcanol. Geotherm. Res.134, 199-221.

Saccorotti, G., Petrosino, S., Bianco, F., Castellano, M., Galluzzo, D., La Rocca, M., Del Pezzo, E., Zaccarelli, L., Cusano, P., 2007. Seismicity associated with the 2004–2006 renewed ground uplift at Campi Flegrei caldera, Italy. Phys. Earth Planet. Inter. 165, 14–24. doi:10.1016/j. pepi.2007.07.006.

Sgattoni, G., Guðmundsson, Ó., Tryggvason, A., Lucchi, F., Einarsson, P., 2015. The 2011 unrest at Katla volcano: location and interpretation of the tremor source. Poster session presented at: 26th IUGG General Assembly 2015. Prague, Czech Republic.

Sigurðsson, O., Zóphoníasson, S., Ísleifsson, E., 2000. Jökulhlaup úr Sólheimajökli 18. júlí 1999 (The jökulhlaup from Sólheimajökull July 18, 1999, in Icelandic with English summary), Jökull 49, 75–80.

Soosalu, H., Jónsdóttir, K., Einarsson, P., 2006. Seismicity crisis at the Katla volcano, Iceland – signs of a cryptodome?. J. Volcanol. Geotherm. Res. 153, 177–186.

Spaans, K., Hreinsdóttir, S., Hooper, A., Ófeigsson, B.G., 2015. Crustal movements due to Iceland's shrinking ice caps mimic magma inflow signal at Katla volcano. Sci. Rep. 5, 10285. doi: 10.1038/srep10285

Sturkell, E., Einarsson, P., Sigmundsson, F., Geirsson, H., Ólafsson, H., Pedersen, R., de Zeeuw-van Dalfsen, E., Linde, A.T., Sacks, S.I., Stefánsson, R., 2006. Volcano geodesy and magma dynamics in Iceland. J. Volcanol. Geotherm. Res. 150, 14–34. doi:10.1016/j.jvolgeores.2005.07.010

Sturkell, E., Einarsson, P., Roberts, M.J., Geirsson, H., Guðmundsson, M.T., Sigmundsson, F., Pinel, V., Guðmundsson, G.B., Olafsson, H., Stefansson, R., 2008. Seismic and geodetic insights into magma accumulation at Katla subglacial volcano, Iceland: 1999 to 2005. J. Geophys. Res. 113, B03212.

Sturkell, E., Einarsson, P., Sigmundsson, F., Hooper, A., Ófeigsson, B. G., Geirsson, H., Ólafsson, H., 2010. Katla and Eyjafjallajökull Volcanoes. In: Schomacker, A., Krüger, J., Kjær, K.H. (Eds). The Mýrdalsjökull icecap, Iceland. Glacial processes, sediments and landforms on an active volcano. Developments in Quaternary Science 13. Elsevier, Amsterdam. ISBN 1571-0866, pp. 5–21.

Thelen, W. A., Allstadt, K., De Angelis, S., Malone, S. D., Moran, S. C., Vidale J., 2013. Shallow repeating seismic events under an alpine glacier at Mount Rainier, Washington, USA. J. Glaciol. 59, 214.

Thorarinsson, S., 1975. Katla og annall Kotlugosa. Arbok Ferdafelags Islands. Ferdafelag Íslands, Reykjavık, 125–149.

Thordarson, T., Miller, D.J., Larsen, G., Self, S., Sigurðsson, H., 2001. New estimates of sulfur degassing and atmospheric mass-loading by the 934 AD Eldgjá eruption, Iceland. J. Volcanol. Geotherm. Res. 108, 33–54.

Weaver, C. S., Malone, S. D., 1979. Seismic evidence for discrete glacier motion at the rock-ice interface. J. Glaciol. 23, 171– 184.

West, M.E., Larsen, C.F., Truffer, M., O'Neel, S., LeBlanc, L., 2010. Glacier microseismicity. Geology, 38(4), 319–322. doi:10.1130/G30606.1

Wiens, D. A., Anandakrishnan, S., Winberry, J.P., King, M.A., 2008. Simultaneous teleseismic and geodetic observations of the stick-slip motion of an Antarctic ice stream. Nature, 453, 770– 774.

Winberry, J.P., Conway, H., Huerta, A.D., Anandakrishnan, S., Aster, R.C., Koutnik, M., Nyblade, A., Wiens, D.A., 2015. Periodic, episodic, and complex behavior of glacial earthquakes. Paper presented at: 26th IUGG General Assembly 2015. Prague, Czech Republic.